\documentclass[aps,prl,groupedaddress,showpacs,superscriptaddress,showkeys,preprint,endfloats]{revtex4}
\usepackage{graphicx}
\usepackage{dcolumn}
\usepackage{bm}
\begin{document}

\title{Restricted Wiedemann-Franz law and vanishing thermoelectric power in one-dimensional conductors}


\author{Marcelo A. Kuroda}
\affiliation{Department of Physics and Beckman Institute, University of Illinois at Urbana-Champaign, Illinois 61801}
\author{Jean-Pierre Leburton}
\email{jleburto@illinois.edu}
\affiliation{Department of Electrical and Computer Engineering and Beckman Institute, University of Illinois at Urbana-Champaign, Illinois 61801}


\date{\today}

\begin{abstract}
In one-dimensional (1D) conductors with linear $E$-$k$ dispersion
(Dirac systems) intrabranch thermalization is favored by elastic
electron-electron interaction in contrast to electron
systems with a nonlinear (parabolic) dispersion. We show that under
external electric fields or thermal gradients the carrier populations of different branches,
treated as Fermi gases, have different temperatures as a
consequence of self-consistent carrier-heat transport. Specifically, in the presence of elastic phonon
scattering, the Wiedemann-Franz law is restricted to each branch with its specific temperature
and is characterized by twice the Lorenz number.  In addition
thermoelectric power vanishes due to electron-hole symmetry, which is validated by experiment.
\end{abstract}


\maketitle

Within the last decades one dimensional (1D) conductors such
as nanowires, nanotubes and molecular chains have become
experimentally accessible \cite{hu1999}.
While electron populations at low temperatures in ideal 1D systems are predicted to behave 
as Tomonaga-Luttinger liquids \cite{solyom1979, voit1995} 
transport experiments in 1D conductors have revealed various
behaviors depending on the temperature range and quality of the
samples. Indeed, at low temperatures  conductance
quantization \cite{yacoby1996}, and signatures of Tomonaga-Luttinger liquid
\cite{bockrath1999,zaitsev2000} and Wigner crystallization
\cite{deshpande2008} have been observed. However, the experimental
realization of such systems is tremendously challenging and still
requires further unambiguous confirmation. 
As temperature is increased, the features of Tomonaga-Luttinger liquids are 
smeared out by thermal broadening and carriers 
behave as Fermi gases. In this regime,
electron transport, ranging from ballistic to diffusive has been
successfully described by semi-classical approaches, such as
Landauer-B\"utikker formalism\cite{park2004} direct solution of the 
Boltzmann equation\cite{kuroda2005, lazzeri2006}, 
and Monte-Carlo
simulations\cite{javey2004}. Paradoxically these approaches often neglect
electron-electron (e-e) interaction as well as self-consistent
heat transport regulating the energy carried by electrons.

In this letter we show that in 1D conductors with linear
energy dispersion (Dirac system) energy and momentum
conservation favors elastic interbranch e-e scattering, in
contrast to 1D systems with nonlinear (parabolic) 
dispersion\cite{leburton1992}. As a consequence,
the fermion populations in different branches are not in thermal
equilibrium, and are characterized by two different temperatures,
even in the lowest electric fields due to the mutual
influence between carrier and heat transport. Our self-consistent
analysis of electro-thermal transport of 1D Dirac systems 
shows that the ratio between thermal and electrical conductivity
is proportional to the branch temperature (Wiedemann-Franz law) with 
a factor equal to twice the Lorenz number.
The thermoelectric power (TEP) in 1D conductors vanishes 
because of electron-hole symmetry.

The Hamiltonian describing 1D Dirac systems can be written as:
\begin{equation}
\hat{H} = v_F \hat{p}_z \hat{\sigma}_y
\end{equation}
where $v_F$ is the Fermi velocity and $\hat{\sigma_y}$  is the $y$ component of the Pauli
spin matrix, which gives the following $E-k$ relationship:
\begin{equation}
E_\pm(k) = \pm \hbar v_F k \label{dispersion}
\end{equation}
for which the $\pm$ sign refers to two different
energy branches and results in a constant density of states..
Here $k$ is the wave vector along the 1D $z$-direction.

\begin{figure}[htbp]
  \leavevmode \centering
    \includegraphics[width=3.25in]{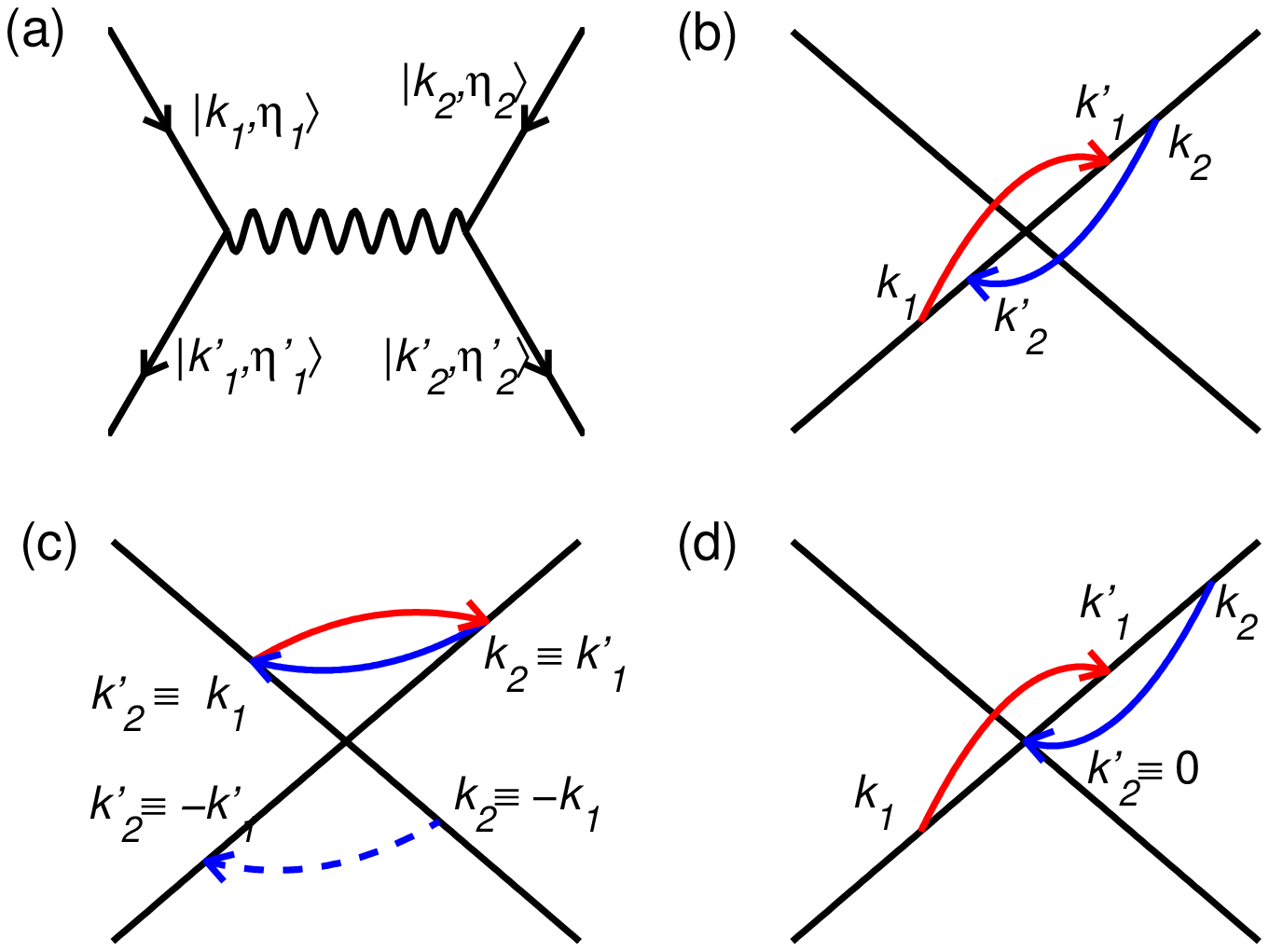}
  \caption{\label{scattering} Scattering diagrams due
to electron-electron interaction in which energy and momentum are
conserved. (a) Feynmann diagram for the elastic e-e scattering.
(b) Intra-intra-branch scattering. (c) Inter-inter-branch
scattering ; (d) Intra-inter-branch scattering}
\end{figure}

For linear band structure elastic e-e collisions are
grouped in three classes of processes, 
i.e.~intra-intra, intra-inter and inter-inter branch
scattering, depending on whether the initial and final states 
remain in the same (intra) branches or change (inter) branches with collisions. 
Hence, we consider scattering
from the initial state $|k_{1},\eta_{1};k_{2},\eta_{2}\rangle$ to
the final state $|k^\prime_{1},{\eta_{1}^\prime};k^\prime_{2},{\eta_{2}^\prime}\rangle$, where $k$ and $\eta$ indicate the wave-vector and sign
of the branch's Fermi velocity ($\eta = +,-$), respectively (Fig.~\ref{scattering}a).
We assume for simplicity that none of these bands is
degenerate, and that there is only one valley, but 
the analysis can be easily extended to degenerate
branches and multiple valleys. We set $k = 0$ at the branches
crossing (Dirac point), and since both 
momentum and energy between initial and final states
are conserved, we get:
\begin{eqnarray}
k_{1}+k_{2}=k_1^\prime+k_2^\prime\label{momcons}\\
E_{\eta_{1}}(k_{1})+E_{\eta_{2}}(k_{2})=
E_{\eta_{1}^\prime}(k_{1})+E_{\eta_{2}^\prime}(k_{2})\label{encons}
\end{eqnarray}

Because of the proportionality between $E$ and $k$ these two equations are
linearly dependent for intra-intra branch transitions 
(i.e.~all the electrons states belong to the same branch). 
For example, for intra-intra branch $|k_{1},+;k_2,+\rangle \rightarrow
|k_{1}^{\prime},+;k_{2}^{\prime},+\rangle$ scattering, multiple
values of $k_{2}$ and $k_{2}^{\prime}$ satisfy energy and momentum
conservation, given an arbitrary pair of values for $k_1$ and
$k_1^\prime$ (Fig.~\ref{scattering}b). 
If any inter-branch transition takes 
place, Eqs.~\ref{momcons} and \ref{encons} become linearly 
independent (as in particle systems with
nonlinear $E$-$k$ dispersion). Fig.~\ref{scattering}c
shows the two possible cases of inter-inter branch
scattering for which the first electron transition is
$|k_{1},-\rangle \rightarrow |k^{\prime}_{1},+\rangle$. The solid
(dashed) arrow to the left corresponds to the exchange (symmetric)
scattering for which $k_{2}=k^\prime_{1}$ and $k^\prime_{2}=k_{1}$
($k_{2}=-k_{1}$ and $k^\prime_{2}=-k^\prime_{1}$) which
is totally inefficient for 
thermalization\cite{leburton1992}. Intra-inter
band transitions occurs if and only if the wave-vector of the
state scattering to the different band is at the Dirac point
(i.e.~$k = 0$)  as shown in Fig.~\ref{scattering}d for
$|k_{1},+;0,-\rangle \rightarrow
|k_{1}^{\prime},+;k_{2}^{\prime},+\rangle$. 
Therefore, the number of collisions
involving only intra-branch scattering processes are roughly $N_k$
(number of $k$-states) larger than the number of inter-intra
or inter-inter branch transitions. Since $N_k$ is proportional to
the number of atoms in the system ($N_k \gg 1$) intra-intra-branch
e-e collisions are much more likely to occur than inter-intra or
inter-inter-branch collisions.

Because intra-intra branch scattering is more efficient
than inter-intra or inter-inter branch scattering, carrier populations
in different branches behave
as independent Fermi gases with specific temperatures. In
the presence of an electric field $F$ an imbalance arises
between carrier populations in different branches
(different quasi-Fermi level $\mu_\eta$)  with a nonzero current flow.
Consequently the distribution functions in each (thermalized) branch reads:
\begin{equation}
f_\eta(E) = \frac{1}{\exp[(E-\mu_\eta)/k_BT_\eta]+1}\label{distfunc}
\end{equation}
where $T_\eta = T_\eta (z)$ is the local electronic temperature of the branch $\eta$.
As the quasi-Fermi levels are far away from the band edges, and
because the thermal broadening is much smaller than the band width,
the carrier densities are independent of the respective temperatures due to the
constant density of states.
Since all the carriers share the same group velocity, the current is
proportional to the quasi-Fermi level difference\cite{kuroda2005}, i.e.
\begin{equation}
I = e v_F (n_+-n_-)=  \frac{g_c e}{\pi \hbar}(\mu_+-\mu_-) =
g_c G_0 \frac{(\mu_+-\mu_-)}{e} \label{current}
\end{equation}
where $G_0 = e^2/(\pi\hbar)$ is the quantum conductance. The factor $g_c$ 
accounts for the band degeneracy (spin degeneracy has already been considered).
In Eq.~\ref{current} the current can be interpreted
as the superposition of the electron current ($I_+$) and
hole current ($I_-$):
\begin{equation}
I_\pm = \pm \frac{g_c e}{\pi\hbar}(\mu_\pm-\mu) = \frac{I}{2}\label{ehcurrent}
\end{equation}
where $\mu\equiv (\mu_+-\mu_-)/2$ is the effective Fermi level of the system.
Similarly the internal energy flow per unit
length for each branch with respect to the effective Fermi level is:
\begin{equation}
U_\pm = \pm\frac{g_c}{\pi \hbar} \int_{-\infty}^{\infty} (E-\mu_\pm)
\left[f_\pm(E)-\Theta(-E+\mu)\right]  dE = \pm\frac{g_c}{\pi \hbar}\left[\frac{\pi^2  (k_BT_\pm)^2}
{6}+\frac{(\mu_+-\mu_-)^2}{8}\right]\label{energyflow},
\end{equation}
where $\Theta(x)$ is the Heaviside step function\cite{inftyapprox}.

In the high (room) temperature semi-classical regime, the distribution function in each branch follows the stationary
Boltzmann transport equation (BTE) which after
making use of Eq.~\ref{dispersion} reads:
\begin{equation}
\pm v_F\left[ \,\partial_z f_\pm(E) + eF  \partial_E
f_\pm(E) \right]=\partial_t f_\pm\Big\arrowvert_{coll} \label{boltzmann}.
\end{equation}
Here $F$ and $\partial_t f_\pm\big|_{coll}$ are
the electric field and the collision integral accounting for
carrier scattering, respectively.
We solve this equation using the method of moments, i.e.~we
multiply the BTE by $(E-\mu)^m$, where $m$ is the moment index,
and integrate over energy, for which we obtain:
\begin{equation}
-e\left(F - \frac{1}{e}\partial_z \mu_\pm\right)  = \pm \frac{2\pi\hbar}{g_c}\mathcal{F}_{0,\pm}^{coll} \label{coll0pm}
\end{equation}
and
\begin{equation}
\pm \partial_zU_\pm \mp \frac{I }{2}\left(F - \frac{1}{e}\partial_z \mu_\eta \right)=\mathcal{F}_{1,\pm}^{coll}\label{coll1pm}
\end{equation}
for the zeroth and first moment equations, respectively.
$ \mathcal{F}_{m,\eta}^{coll}\equiv \frac{g_c}{2\pi\hbar v_F}\int_{-\infty}^{\infty}
(E-\mu) ^m\partial_t f_\eta\big|_{coll} dE \label{collint}$ is the
generalized $m$th moment of the collision integral for each branch.
The left-hand side of Eq.~\ref{coll0pm} is the measured
effective field ($F' \equiv F - \partial_z\mu_\pm/e$) \cite{ziman}, and we
note that in general the expressions for $\mathcal{F}_{m,\eta}$
depend on both the local temperatures and current level (or
difference between quasi-Fermi levels). This dependence couples the equations
describing the heat flow (Eq.~\ref{coll1pm}) for each carrier population.
We solve both equations for the field and the temperatures profiles by
using the current level as a parameter.

In the case of 1D metals, the main scattering process
(other than e-e interaction)
is scattering with phonons. We compute the collision integral for this mechanism  as:
\begin{eqnarray}
\frac{\partial f_\eta}{\partial t}\Big\arrowvert_{coll} =
\sum_{q,\eta'} \left\{R_{em}(q) \left[f_{\eta'}(\epsilon_+)\left(1-
f_\eta(E)\right)-f_\eta(E)\left(1-
f_{\eta'}(\epsilon_-)\right)\right]\right.\nonumber\\
\left.R_{ab}(q) \left[f_{\eta'}(\epsilon_-)(1-
f_\eta(E)-f_\eta(E)(1 - f_{\eta'}(\epsilon_+)\right] \right\}
\end{eqnarray}
where $\epsilon_\pm= E\pm\hbar\omega_q$ and
$q$ labels the phonon wave-vector. The prefactor $R_{em}$ ($R_{ab}$)
is the phonon emission (absorption) rate.
Here we focus on acoustic phonon scattering  ($\omega_q = v_s |q|$),
which we assume to be elastic (i.e.~$v_s/v_F \ll 1$), and for which
only the longitudinal modes contribute to the integral. Since phonon energies
are much smaller than the lattice temperature $T_L$ ($\hbar\omega_{q}\ll k_BT_L$) and using
the deformation potential approximation, we set
$R_{em}=R_{ab}\propto T_L$ \cite{hess}. We then define
the mean free path $\lambda_{ac}\equiv v_F/R_{em}$.

By subtracting and adding the 0th moments of the BTE
(Eq.~\ref{coll0pm}) for the two carrier branches, we obtain:
\begin{eqnarray}
\partial_z (\mu_+-\mu_-) =0 \label{curr_cons}
\end{eqnarray}
and
\begin{eqnarray}
I = g_c G_0 \lambda_{ac} F'\label{odeeFpm}
\end{eqnarray}
respectively. Eq.~\ref{curr_cons} expresses the local current
conservation in the system ($ I \propto \mu_+-\mu_-$). 
Eq.~\ref{odeeFpm} is nothing but Ohm's law for which the linear
conductivity is inversely proportional to the lattice temperature
($\lambda_{ac}\propto T_L^{-1}$) but independent of the carrier
temperatures $T_\pm$. 

Using Eq.~\ref{ehcurrent} we define the branch electrical conductivity
as $\sigma_\eta \equiv g_c G_0 \lambda_{ac}/2 $.
We emphasize that owing to the electron-hole
symmetry Eq.~\ref{odeeFpm} does not depend on the
thermal gradient. Therefore, the thermoelectric power 
(TEP) in 1D (Dirac) conductors vanishes, which is in agreement 
with recent experiments on metallic carbon nanotubes that exhibit
much smaller TEP than that of semiconducting
carbon nanotubes \cite{small2003}.

In the presence of acoustic phonon scattering the 1st moment
equation (Eq.~\ref{coll1pm}) for each branch reads:
\begin{equation}
\partial_z U_\pm =
\frac{I F'}{2}\mp\frac{1}{\lambda_{ac}}
\left(U_++U_-\right)\label{odeTpm}
\end{equation}
Using Eqs.~\ref{energyflow} and \ref{curr_cons}, the LHS of
Eq.~\ref{odeTpm} is proportional to $\partial_z T_\pm$.
This variation in the carrier temperature profiles is attributed
to the Joule heating (first term on the RHS) 
and the inter-branch carrier scattering due to
the carriers thermal imbalance (second term). 

Combining Eq.~\ref{odeTpm} for the different branches we establish
the heat flow conservation:
\begin{equation}
\partial_z(U_++U_-) =I F' \label{energy_cons}
\end{equation}
which shows that all the heat is transported by the electrons
and results in an inhomogeneous temperature profile. This is consistent
with our approximation of elastic phonon scattering for which no
energy gained by the carriers from the
external field is transferred to the lattice
(i.e.~the lattice temperature remains constant).
The energy production/dissipation in the system
couples the two carrier temperatures $T_\pm$ (Eq.~\ref{odeTpm}).
Furthermore, Eq.~\ref{energy_cons} validates the assumption
of the two temperature model for the electronic population in each branch.
Indeed, if $T_+(z) = T_-(z)$ the LHS of Eq.~\ref{energy_cons} vanishes,
which is inconsistent with a non-zero current.

Substituting Eq.~\ref{energyflow} 
into Eq.~\ref{energy_cons} we obtain for the heat flow:
\begin{equation}
\mp\frac{\lambda_{ac}g_c \pi k_B^2T_\pm}{3\hbar} \partial_zT_\pm = U_++U_-\mp \frac{\lambda_{ac} I F'}{2}
\end{equation}
for which the pre-factor  in the temperature gradient is the carrier thermal conductivity:
\begin{equation}
\kappa_\eta=\frac{\lambda_{ac}g_c \pi k_B^2T_\eta}{3\hbar} =
\lambda_{ac} g_c G_{th}
\end{equation}
where $G_{th}$ is the quantized thermal conductance \cite{rego1997}.
The ratio between thermal and electrical conductivity for each branch:
\begin{equation}
\frac{\kappa_\eta}{\sigma_\eta T_\eta} = \frac{2\pi^2k_B^2}{3 e^2}
\end{equation}
obeys Wiedemann-Franz law with their specific carrier temperature 
and a proportionality factor that is twice the Lorenz number (for 2 and 3 dimensions) \cite{ziman}.

\begin{figure}[htpb]
\includegraphics[width=3.25in]{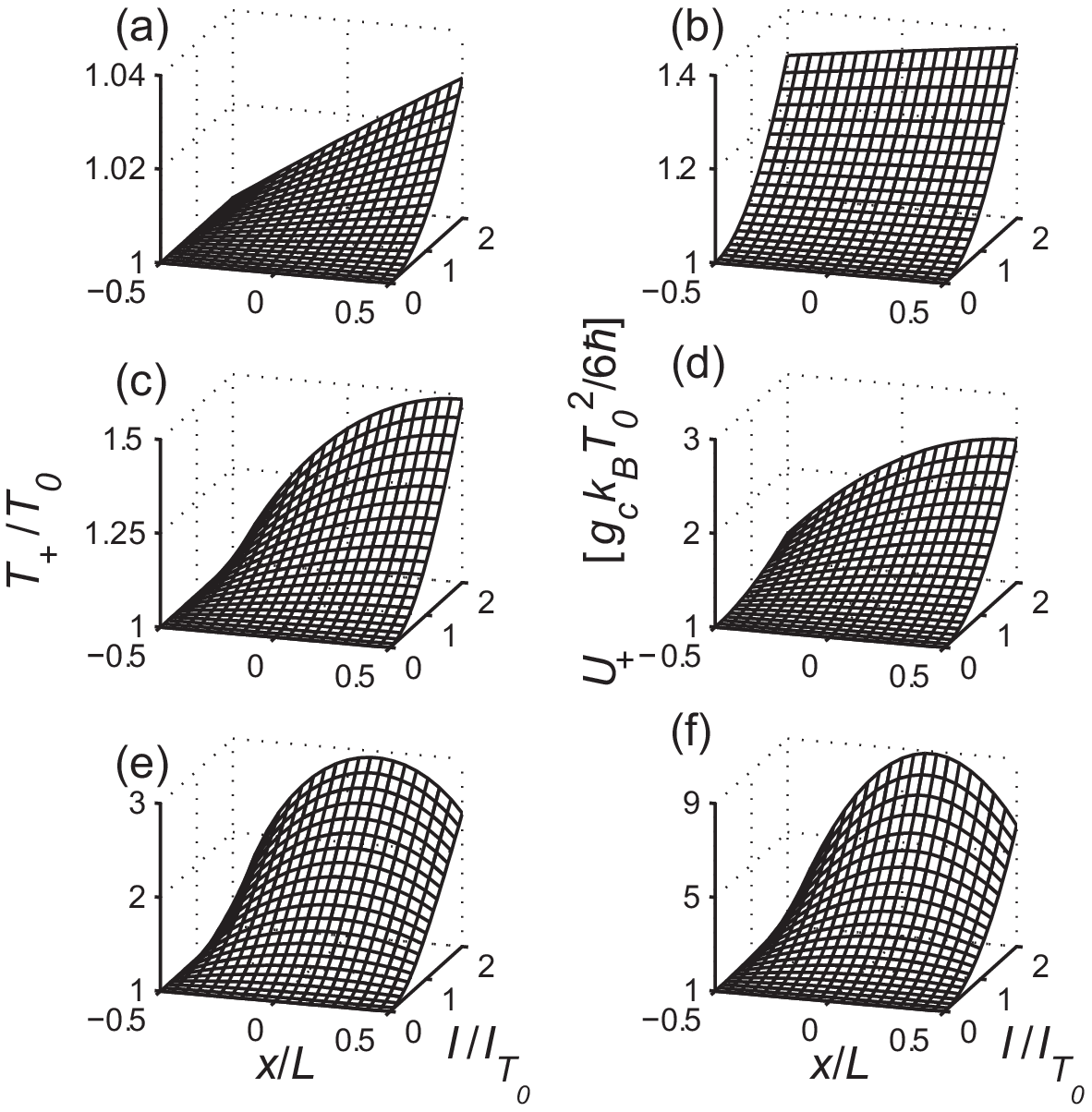}
\caption{Temperature profiles ($T_+$) and heat flow 
$U_+$ for carriers with positive Fermi velocity
as a function of the current for: $\lambda_{ac} = 20 L$ (a,b),
 $\lambda_{ac} = L$ (c,d), and  $\lambda_{ac} = 0.25 L$ (e,f).}
\label{IvsF}
\end{figure}

In the case of isothermal lattices, $\lambda_{ac}$ and $F$ are constant,
and by assuming ideal contacts (i.e.~completely absorbing and thermalizing) $T_\pm(\mp L/2) = T_0$ ,
the solution of Eq.~\ref{odeTpm} is:
\begin{eqnarray}
T_\pm(z)=T_0\sqrt{1 +\frac{3}{\pi^2}\left[1 +
\frac{1\mp2z/L}{2\lambda_{ac}/L}\right]
\frac{1\pm2z/L}{2\lambda_{ac}/L}\left(\frac{I}{I_{T_0}}\right)^2}\label{Tpmprof}
\end{eqnarray}
for which $I_{T_0}\equiv (g_c e k_B T_0)/(\pi \hbar)$ 
is the current associated to the lattice temperature.
The temperature of the carriers with positive (negative) Fermi velocity has a maximum at
$z =\lambda_{ac}$ ($z=-\lambda_{ac}$).
In the quasi-ballistic limit, $\lambda_{ac}\gg L$, 
the carrier temperature profiles $T_\pm^{QB}(z)$ have a linear $z$-dependence:
\begin{equation}
\frac{T_\pm^{QB}(z)}{T_0} = 1+\frac{3 L}{4\pi^2 \lambda_{ac}}\left(\frac{I}{I_{T_0}}\right)^2
\left(1\pm\frac{2z}{L}\right)
\end{equation}
In this regime, the temperature difference $T_\pm-T_0$ is small even
for $I \sim I_{T_0}$ because of the ratio $L/\lambda_{ac} \ll 1$.
In the diffusive regime ($\lambda_{ac}\ll L$), the carrier temperature profiles 
have a parabolic shape and the maximum temperature difference 
is of the same order as $T_0$
even for small current levels. 
In Fig.~\ref{IvsF} we display the temperature and heat flow profiles 
corresponding to carrier populations with positive Fermi velocity 
($T_+(z,I)$ and $U_+(z,I)$) for ratios $L/\lambda_{ac} = 0.05$, 1 and 4, respectively.
The electron-hole symmetry in this case is expressed by
the fact that $T_+(z) = T_-(-z)$ and $U_+(z) = -U_-(-z)$.

If a lattice temperature gradient exists along the conductor
$T_L(z) = T_0-\Delta T\, z/L)$, the
carrier temperature profiles are:
\begin{eqnarray}
T_\pm(z)=\sqrt{\left(T_0\pm\frac{\Delta T}{2}\right)^2+
\frac{L}{L+\lambda_0}\left[\left(T_0-\Delta T \frac{z}{L}\right)^2-
\left(T_0\pm\frac{\Delta T}{2}\right)^2\right]}\label{Tprof}
\end{eqnarray}
for which we have assumed $\lambda_{ac}(z) T_L(z) = \lambda_0 T_0 $ 
(Eq.~\ref{odeeFpm}) and the symmetric 
boundary conditions $T_L(-L/2) = T_+(-L/2)$ and
$T_L(L/2) = T_-(L/2)$. For $\Delta T \ll T_0$,
Eq.~\ref{Tprof} becomes:
\begin{eqnarray}
T_\pm(z)=T_0-\frac{\Delta T}{2}\frac{(2 z \mp \lambda_0)}{(L+\lambda_0)}
\end{eqnarray}
and the carrier temperature difference reads,
\begin{equation}
T_+(z)-T_-(z)=\Delta T\frac{\lambda_{ac}}{L+\lambda_{ac}}.
\end{equation}
In the case of ballistic transport ($\lambda_{ac}/{L} \gg 1$)
the temperature difference remains $T_+-T_- = \Delta T$ along the
conductor, while for diffusive transport ($\lambda_{ac}/{L} \ll 1$), the
temperature difference between branches is negligible.

It is important to emphasize that Eqs.~\ref{coll0pm} and \ref{coll1pm}
are valid both in low and high field regimes. However, here we
only consider elastic scattering and thereby the results are valid in the
linear regime (i.e.~low current levels and $\hbar\omega_q \ll k_B T_L$).
At high current levels, other scattering processes (e.g. optical phonon scattering)
have to be included in the collision integrals.

In conclusion because of the effective intrabranch carrier thermalization
in the high temperature regime electron populations 
in the different branches of 1D Dirac conductors
behave as independent Fermi gases (with their respective temperatures) out of
thermal equilibrium as a consequence of the electro-thermal flow.
In the presence of acoustic phonon scattering, the carrier population
in each energy branch follows the Wiedemann-Franz law characterized by twice the 
Lorenz number. The TEP coefficient in
1D conductors vanishes as a result of the electron-hole symmetry.

\bibliography{report}

\begin{thebibliography}{17}
\bibitem[{inf()}]{inftyapprox}
\bibinfo{note}{The Heaviside function is introduced to eliminate the artificial
  divergence arising from the extension of the bottom of the conduction bands
  to infinity in the integration.}
\end{thebibliography}
\end{document}